\documentclass[aps,prl,reprint,amsmath,raggedbottom]{revtex4-1}
\usepackage{graphicx}
\usepackage[colorlinks,allcolors=blue]{hyperref}
\usepackage[figure,figure*]{hypcap}

\newcommand{\pref}[2]{\hyperref[#1]{\ref{#1}(#2)}}

\newcommand{\eqpref}[1]{\hyperref[#1]{(\ref{#1})}}

\renewcommand{\section}[1]{{\vskip\baselineskip\fontfamily{cmss}\fontseries{bx}\selectfont\noindent \pdfbookmark[1]{#1}{#1} #1}}
\renewcommand{\subsection}[1]{\vskip\baselineskip \pdfbookmark[2]{#1}{#1} \emph{#1}}

\newcommand{\ket}[1]{|#1\rangle}

\newcommand{\jzsq}[0]{J_z^2}
\newcommand{\jsn}[2]{$\ket{J=#1,m_J=#2}_z$}

\newcommand{\css}[2]{$\ket{\theta=#1,\phi=#2}_{\textrm{CSS}}$}

\begin{document}

\title{Exploring quantum signatures of chaos on a Floquet synthetic lattice}
\author{Eric J. Meier}
\thanks{These authors contributed equally to this work}
\author{Jackson Ang'ong'a}
\thanks{These authors contributed equally to this work}
\author{Fangzhao Alex An}
\author{Bryce Gadway}
\email{bgadway@illinois.edu}
\affiliation{Department of Physics, University of Illinois at Urbana-Champaign, Urbana, IL 61801-3080, USA}
\date{\today}

\begin{abstract}
Ergodicity and chaos play an integral role in the dynamical behavior of many-particle systems and are crucial to the formulation of statistical mechanics. Still, a general understanding of how randomness and chaos emerge in the dynamical evolution of closed quantum systems remains elusive. Here, we develop an experimental platform for the realization of canonical quantum chaotic Hamiltonians based on quantum simulation with synthetic lattices. We map the angular momentum projection states of an effective quantum spin onto the linear momentum states of a $^{87}$Rb Bose-Einstein condensate, which can alternatively be viewed as lattice sites in a synthetic dimension. This synthetic lattice, with local and dynamical control of tight-binding lattice parameters, enables new capabilities related to the experimental study of quantum chaos. In particular, the capabilities of our system let us tune the effective size of our spin, allowing us to illustrate how classical chaos can emerge from a discrete quantum system. Moreover, spectroscopic control over our synthetic lattice allows us to explore unique aspects of our spin's dynamics by measuring the out-of-time-ordered correlation function, and enables future investigations into entirely new classes of chaotic systems.
\end{abstract}

\maketitle

\pdfbookmark[1]{Introduction}{Introduction}

The divergent behavior of quantum and classical systems is most apparent in their nonlinear dynamical response to a periodic drive~\cite{chaosbook}. While driven classical systems can play host to truly chaotic behavior, including the loss of information about initial conditions, it is expected that such memory loss will not occur in closed and bounded quantum systems~\cite{HoggHuberman-Recurrence}. Over the past few decades a number of experimental systems have illustrated this stark contrast between the nonlinear dynamics of classical and quantum systems, e.g. the spectra of atoms in applied electromagnetic fields~\cite{GalvezDrivenHydrogen,Kleppner-Chaos-Lithium}, the response of cold matter waves to time-periodic optical lattices~\cite{Moore-kickedrotor,Hensinger-Chaos,Steck-Chaos,Garreau201731,Gadway-QtoC}, and the scattering of complex atoms and molecules in an applied field~\cite{Bohn-Molecule-Chaos,Frisch-Chaos}.

The kicked top model, in which the symmetry of a precessing spin is broken by a series of nonlinear ``kicks''~\cite{chaosbook}, is one of the most paradigmatic systems giving rise to chaotic behavior. The correspondence between the nonlinear dynamics of classical and quantum systems has been explored through several experimental realizations~\cite{Jessen-csspin,Martinis-quditspin,Oberthaler-Chaos-2017} of quantum kicked top models, where the spin is quantized with a finite angular momentum value $J$. In a pioneering exploration of chaotic phenomena in quantum systems, Ref.~\cite{Jessen-csspin} studied the dynamics of the ground hyperfine manifold ($F = 4$) of thermal cesium atoms. The atoms were subjected to a continuous nonlinear twist realized through a state-dependent light shift of the magnetic sublevels ($m_F$) and a periodic linear kick given by a transverse magnetic field. While such studies could be extended to slightly smaller or larger spins with different atomic species, a more flexible approach to designing effective spins with tunable size has recently been realized. Using spectrally-resolved addressing of transitions in a multi-level superconducting qudit, Ref.~\cite{Martinis-quditspin} demonstrated the engineering of artificial spin-$J$ systems and control over linear rotations.

\begin{figure}%
\includegraphics[width=243pt]{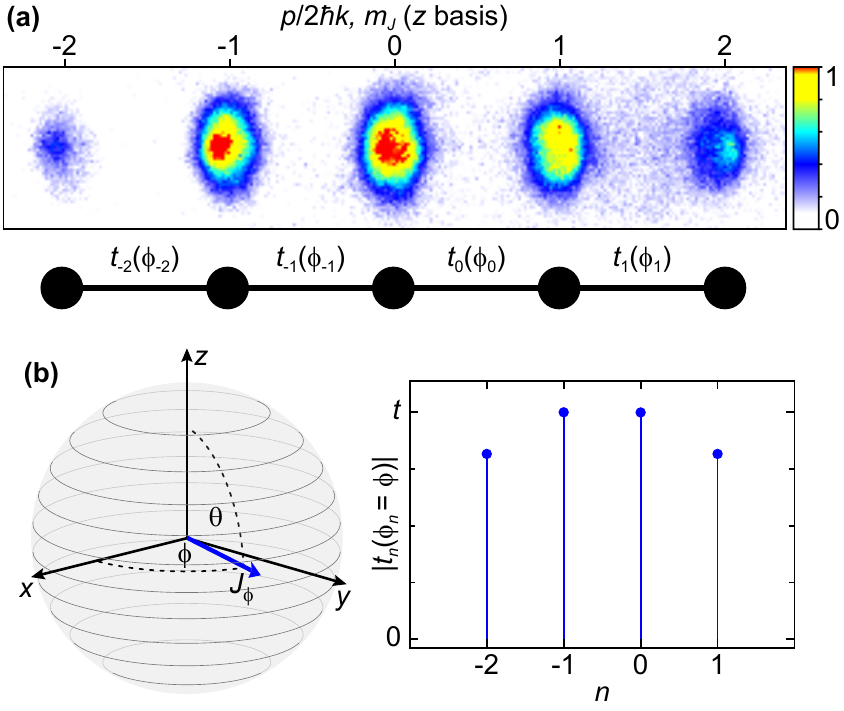}%
\caption{Experimental scheme. \textbf{(a)} Absorption image (top) and cartoon (bottom) depicting a $J=2$ lattice where the lattice sites represent the angular momentum sublevels $m_J$. \textbf{(b)} Arbitrary torque vector on the equator of the collective Bloch sphere (left) emulated in this system through the tunneling links $|t_n(\phi_n=\phi)|$ (right).}%
\label{introfig}%
\end{figure}

Here, in the spirit of creating effective spins through coherent control, we engineer a highly-tunable momentum-space lattice~\cite{MSL-expt,MSL-theory} with full control over the tunneling and site-energy landscapes. In our approach, the sites of a $(2J+1)$-site lattice play the role of angular momentum sublevels $m_J \in \{-J,J\}$ (see Fig.~\pref{introfig}{a}), enabling natural control over the size of the spin $J$. This simple control over $J$ allows us to study the crossover from a highly quantum regime (small $J$), where chaotic behavior is mostly suppressed, to the nearly classical limit (moderate to large $J$), where chaotic~behavior is predicted to emerge.

\subsection{Kicked top model.} The dynamics of the kicked top system are captured by the time-dependent Floquet Hamiltonian
\begin{equation}
H(\tau) =  \frac{\rho}{T} J_x + \frac{\kappa}{2J}\jzsq\sum_{N}\delta(\tau - N T),
\label{KTHam}
\end{equation}
where the first term represents continuous rotation about the $x$ axis at a rate $\rho/T$ and the second describes a train (with period $T$) of effectively instantaneous torsional $\jzsq$ kicks of strength $\kappa$, with $N$ the kick number and $\tau$ the time variable. In the classical limit, symmetry-breaking by the $\jzsq$ kicks gives rise to chaotic dynamics for certain initial orientations of the spin, with islands of stability in phase space for moderate nonlinear coupling. As $\kappa$ is increased, the onset of global chaos leads to the loss of all stable, regular trajectories of the spin. In the limit of small $J$, the lack of well-defined spin orientations due to quantum uncertainty results in a general insensitivity to initial conditions and chaotic behavior.

Connections between classical chaos and the generation of quantum entanglement~\cite{zurek-entropy,entropy-expt} add further interest to the interplay between classical and quantum dynamics. For quantum kicked top dynamics in which the spin-$J$ object represents the collective spin of many interacting spin-$1/2$ particles (e.g., in atomic condensates with a spin degree of freedom~\cite{Oberthaler-Chaos-2017}) scenarios leading to classical chaos can generate quantum correlations and metrologically useful spin squeezing~\cite{kitagawa-spinsqueezing}. Starting from a coherent spin state (CSS) $\ket{\theta,\phi}$ (where all of the spin-$1/2$ particles are in the same superposition state $\cos(\theta/2)|$$\uparrow\rangle+e^{i\phi}\sin(\theta/2)|$$\downarrow\rangle$) the states of the individual particles become entangled and the many-body state becomes non-separable under the evolution of Eq.~\ref{KTHam}. The direct measurement of multi-particle correlations generated by collective kicked top dynamics has recently been achieved for the small $J$ limit, in a system of superconducting qubits with engineered collective interactions~\cite{Neill-Martinis-Chaos}.

Here, instead of studying the collective spin of many interacting spin-$1/2$ particles, we directly mimic the dynamics of a single spin-$J$ quantum object. To successfully explore quantum chaos in this system, we must be able to engineer an effective spin system, realize the kicked top Hamiltonian of Eq.~\ref{KTHam}, accurately prepare initial states of the spin, and measure the final state of the spin after some dynamical evolution. In the following sections, we describe how we achieve these tasks using momentum-space lattice techniques.

\begin{figure*}%
\includegraphics[width=336pt]{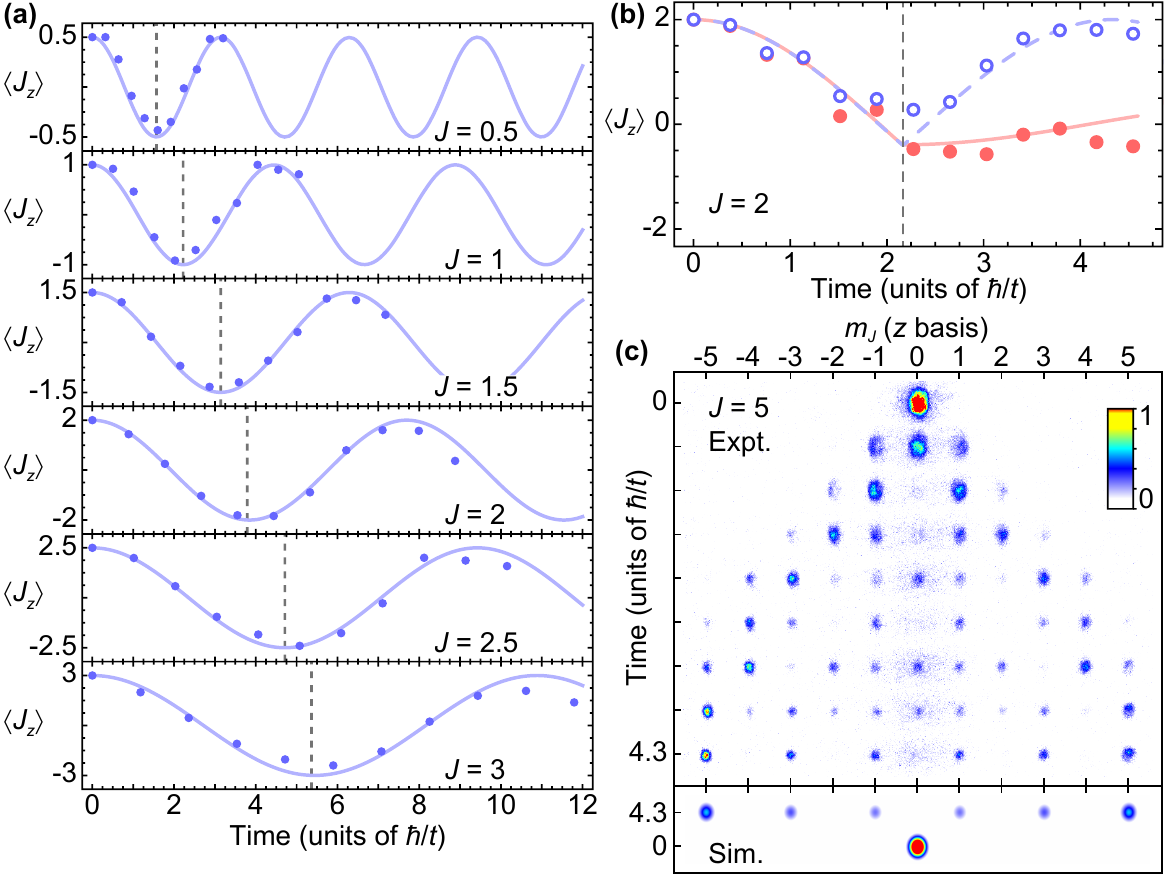}%
\caption{Demonstrations of linear rotations. \textbf{(a)} Evolution of $\langle J_z \rangle$ for several spin sizes starting in \css{0}{0} and evolving under a $J_x$ operator. The solid blue lines are results from simulations of Eq.~\ref{basictb} with no free parameters and the dashed gray lines show the theoretical collective $\pi$ pulse times. \textbf{(b)} Expectation value $\langle J_z \rangle$ for a collective spin-2 state evolving under a $J_x$ operation until the gray dashed line at $\approx$~2.2~$\hbar/t$. At this time, the operation is switched to either $J_{-y}$ (red dots and solid red theory line) or $J_{-x}$ (open blue dots and dashed blue theory line). \textbf{(c)} (top) Experimental absorption images showing the evolution of a $J=5$ spin starting in \jsn{5}{0} evolving under a $-J_y$ operator. (bottom) Simulated absorption images showing the final atomic distribution and the initial state \jsn{5}{0}. All error bars are one standard error of the mean.}%
\label{JzvsJfig}%
\end{figure*}

\subsection{The momentum-space lattice as an artificial spin.} We engineer an artificial spin and realize dynamics governed by Eq.~\ref{KTHam} by coupling many discrete momentum states in a controlled and time-dependent fashion. When considering the $z$-basis projections of the spin, i.e. the $m_J$ sublevels, as sites of a synthetic lattice of discrete quantum states, the two terms of Eq.~\ref{KTHam} allow for a simple realization in terms of lattice dynamics. The $J_x$ rotation can be viewed as a kinetic evolution enabled by tunneling between nearest-neighbor sites. The nonlinear $\jzsq$ kicks are simply instantaneous site-dependent phase shifts, or alternatively represent evolution without tunneling for a fixed time in a quadratic potential of site energies. We realize these elementary processes in a one-dimensional momentum-space lattice~\cite{MSL-expt,MSL-theory} populated by atoms from a $^{87}\textrm{Rb}$ Bose--Einstein condensate, as depicted in Fig.~\pref{introfig}{a}.

Our momentum-space lattice is created from two counter-propagating laser beams with a nearly common wavelength $\lambda = 1064$~nm and wavevector $k = 2\pi/\lambda$. One of the beams has only a single frequency component, while its counter-propagating partner contains multiple discrete frequency components. Initially at rest, the atoms transition between discretized momentum states $p_n = 2n\hbar k$ (separated by twice the photon recoil momentum) by exchanging photons between the two laser beams. That is, the atoms undergo a Bragg diffraction process where they are virtually excited by a photon from one laser beam and then undergo stimulated emission of a photon into the counter-propagating beam, resulting in a $\pm 2\hbar k$ momentum change. The frequencies of the many components of the multi-frequency laser are chosen to match different two-photon Bragg resonance conditions, creating a set of resonantly-connected momentum states that serve as the sites of the momentum-space lattice. By careful tuning of the number, frequency, amplitude, and phase of the components of the multi-frequency beam, we exert full control over the number of sites, site energies, tunneling strengths, and tunneling phases in our lattice, respectively~\footnote{For more detail on the theory and experimental realization of the momentum space lattice see the Methods Summary section of this paper as well as Refs.~\cite{MSL-theory,MSL-expt}}. During an 18~ms time-of-flight expansion period at the end of every experimental cycle, the atoms at different sites of the lattice naturally separate from each other according to their momenta, which allows us to perform site-resolved measurements through standard absorption imaging.

\subsection{Linear spin operators: rotations.} The linear spin operator $J_x$ ($J_y$) can be visualized as the rotation of a given spin state about a torque vector lying on the equator $(\theta = \pi/2)$ of the collective Bloch sphere. This collective Bloch sphere picture also affords a simple visualization of a CSS $\ket{\theta,\phi}$, where the collective spin is oriented along the polar and azimuthal angles $\theta$ and $\phi$, respectively. Alternatively, $J_x$ and $J_y$ can be understood as the matrix representations of the magnetic dipole operator between different $\ket{J,m_J}$ states in a transverse magnetic field.

In order to implement generic rotations about equatorial torque vectors pointing along any azimuthal angle $\phi$, i.e. $J_\phi = J_x \cos(\phi) + J_y \sin(\phi)$, we tailor the tunneling amplitudes and phases between neighboring lattice sites as depicted in Fig.~\pref{introfig}{b}. We introduce tunneling terms $t_n(\phi_n)$ linking lattice site $n$ to site $n+1$ with tunneling phase $\phi_n$, taking the form of the matrix elements of the desired collective spin operator:
\begin{equation}
t_n(\phi_n) = A \sqrt{J(J+1)-n(n+1)}e^{i\phi_n}.
\end{equation}
Here, $n \in \{-J,J-1\}$ is the site index and $A$ is a constant with units of energy related to the tunneling rate. This tunneling function has a maximum amplitude at the center of the $m_J$ manifold, which we label $t$ for convenience (see Fig.~\pref{introfig}{b}). Using these tunneling links we simulate the tight-binding Hamiltonian
\begin{equation}
H_{\textrm{tb}}(\phi_n)=\sum_{n = -J}^{J-1} \left(t_n(\phi_n) c_{n+1}^\dagger c_n +\textrm{h.c.}\right),
\label{basictb}
\end{equation}
where $c_n^\dagger (c_n)$ creates (annihilates) a particle at site $n$. The tunneling phase $\phi_n$ determines the direction of the effective torque vector in the $x$-$y$ plane, where $J_x$ and $J_y$ relate to $H_\textrm{tb}(\phi_n=0)$ and $H_\textrm{tb}(\phi_n=\pi/2)$, respectively.

Figure~\ref{JzvsJfig} summarizes our ability to perform these linear, equatorial spin rotations. Beginning from stretched states ($\ket{J, m_J=J}$), we monitor the $z$-axis projection of the spin evolving under a $J_x$ operator for several values of $J$ (Fig.~\pref{JzvsJfig}{a}). The observed dynamics are in good agreement with theory, with the observed times of spin-inversion (collective $\pi$-pulse times) matching well with theory predictions (dashed lines) for varying $J$~\footnote{in units of $\hbar/t$, this time should scale as $(\pi/2)\sqrt{J(J+1)}$ for integer $J$ and $(\pi/2)(J+1/2)$ for half-integer $J$}.

We further illustrate our phase- and time-dependent control over spin operations in Fig.~\pref{JzvsJfig}{b}. For an initial spin state $|J=2,m_J = 2\rangle$, we first apply a $J_x$ rotation for a time corresponding to a $\pi/2$ pulse. We then modify our tunneling parameters to instantly change the direction of the effective torque vector. For a complete inversion of the torque vector to $-J_{x}$ (evolution under $H_\textrm{tb}(\pi)$), we find that the dynamics of the spin reverse towards the initial state (open blue circles). If we instead shift the torque vector to $-J_{y}$ (evolution under $H_\textrm{tb}(-\pi/2)$), we find that the dynamics essentially cease (red filled circles), since the spin is aligned along the new torque vector. Continued evolution of the spin as seen in Fig.~\pref{JzvsJfig}{b} is due to the spin rotating further than desired before switching the torque vector.

\subsection{State preparation.} As demonstrated in Fig.~\pref{JzvsJfig}{a,b}, we are able to prepare our spin in the stretched state $|J,m_J = J\rangle$ by a simple definition of the synthetic lattice site index with respect to the discrete momentum values ($m_J = J + p/2\hbar k$), and a corresponding choice of the applied Bragg resonance frequencies. This state can be visualized as the collective spin of many spin-$1/2$ particles, all pointing up. We furthermore initiate the spin in any state with well-defined angular momentum in the $z$ basis $|J,m_J\rangle$ by simply defining the corresponding site of our synthetic lattice to match our zero-momentum condensate atoms. The initial states $m_J \neq \pm J$ are effectively squeezed with respect to the operators $J_x$, $J_y$, and $J_z$, granting us the unique ability to explore correlated states evolving under a classically chaotic Hamiltonian. For example, Fig.~\pref{JzvsJfig}{c} shows the evolution of the state $|J=5,m_J = 0\rangle$ under a $-J_y$ spin rotation. This pure angular momentum state displays interesting dynamics as it is rotated. For example, when measured in the $y$ basis after a $\pi/2$ rotation (an evolution time of $\sim 4.3$~$\hbar / t$) a highly-modulated $m_J$ distribution is observed, in excellent agreement with a direct numerical simulation (bottom plot).

\begin{figure}%
\includegraphics[width=243pt]{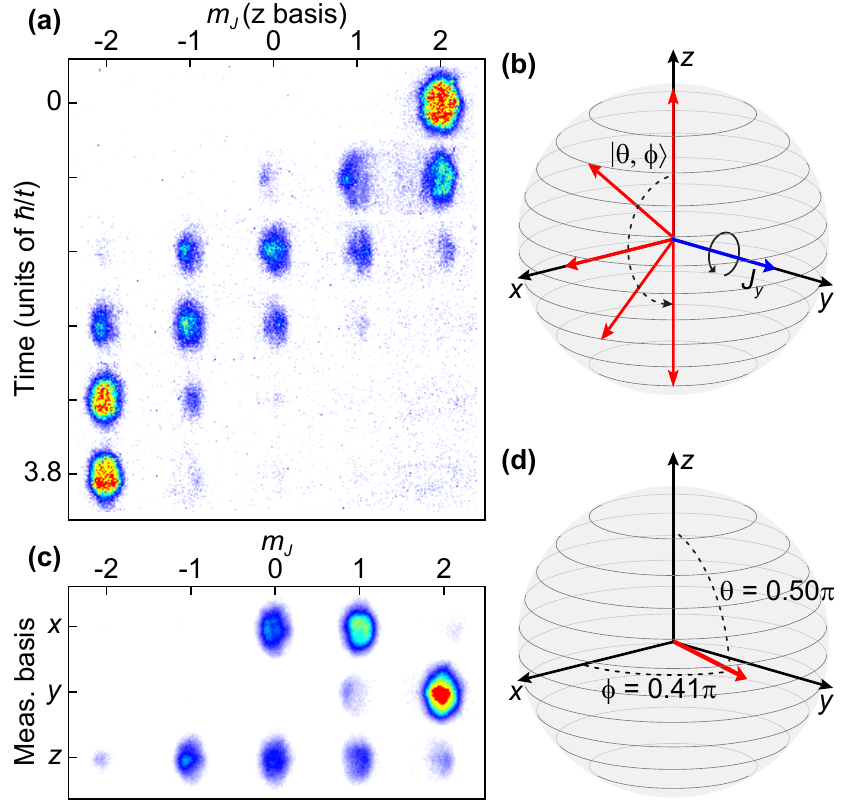}%
\caption{State preparation and measurement. \textbf{(a)} Absorption images (in the $z$ basis) of a $J=2$ spin rotating from \css{0}{0} to \css{\pi}{0} under a $J_y$ operator. \textbf{(b)} Collective Bloch sphere representation of the state rotation shown in (a). The state vector is depicted by the red arrows and the $J_y$ operator by the blue arrow. \textbf{(c)} Images (averaged over many shots) of a $J=2$ spin in the state \css{0.50\pi}{0.41\pi} as measured along the $x$, $y$, and $z$ bases. \textbf{(d)} Collective Bloch sphere depiction of the measured vector shown in (c).}%
\label{stateprepfig}%
\end{figure}

We may also prepare CSSs $\ket{\theta,\phi}$ relating to a collection of $2J$ identical spin-$1/2$ particles each in the superposition state $\cos(\theta/2)\ket{$$\uparrow} + e^{i\phi}\sin(\theta/2)\ket{$$\downarrow}$. To prepare a CSS, we start by initializing our condensate at the north pole of the collective Bloch sphere, i.e. $m_J=J$. Since this state is equivalent to \css{0}{0}, we can apply a collective, coherent rotation to move it to any CSS on the collective Bloch sphere as shown in Fig.~\pref{stateprepfig}{a,b}. In the following experiments we create an arbitrary CSS with parameters $\ket{\theta_i,\phi_i}$ by applying tunneling links $t_n(\phi_i+\pi/2)$ for a time corresponding to a $\theta_i$ pulse. This takes the CSS at the north pole and brings it down along a constant azimuthal angle $\phi_i$ to a polar angle $\theta_i$. Figure~\pref{stateprepfig}{a,b} shows rotation of the state vector (red arrows) about a $J_y$ operator (blue arrow) from \css{0}{0} to \css{\pi}{0}.

\subsection{State measurement.} One nice feature of momentum-space lattices is the ability to measure population at each lattice site directly through simple time-of-flight absorption imaging. In the context of studying the dynamics of an effective spin-$J$ particle on a $(2J + 1)$-site lattice, this relates to directly measuring the $m_J$ state distribution in the $z$ basis. Further information about the quantum state of this artificial spin can be accessed by measuring the spin projection along alternative spin axes, i.e. along the $J_x$ and $J_y$ spin directions. We perform these measurements, related to the coherences between $z$-basis states, by applying a linear rotation about a chosen torque vector prior to $z$-basis imaging. That is, to measure along the $x(y)$ axis we apply a $-J_y(J_{x})$ rotation for a time corresponding to a $\pi/2$ pulse prior to time-of-flight absorption imaging. Figure~\pref{stateprepfig}{c} shows a particular CSS as measured in the $x$, $y$, and $z$ spin bases, while Fig.~\pref{stateprepfig}{d} shows the reconstructed state vector on the collective Bloch sphere, relating to mean-values $\langle J_x \rangle$, $\langle J_y \rangle$, and $\langle J_z \rangle$ of this separable CSS~\footnote{a full reconstruction of the density matrix for a spin of size $J$ be accomplished through measurement in an appropriate choice of $(2J+1)^2 - 1$ bases, allowing for direct visualization in terms of the experimental Husimi-Q distribution~\cite{Jessen-csspin}}.

\begin{figure*}%
\includegraphics[width=480pt]{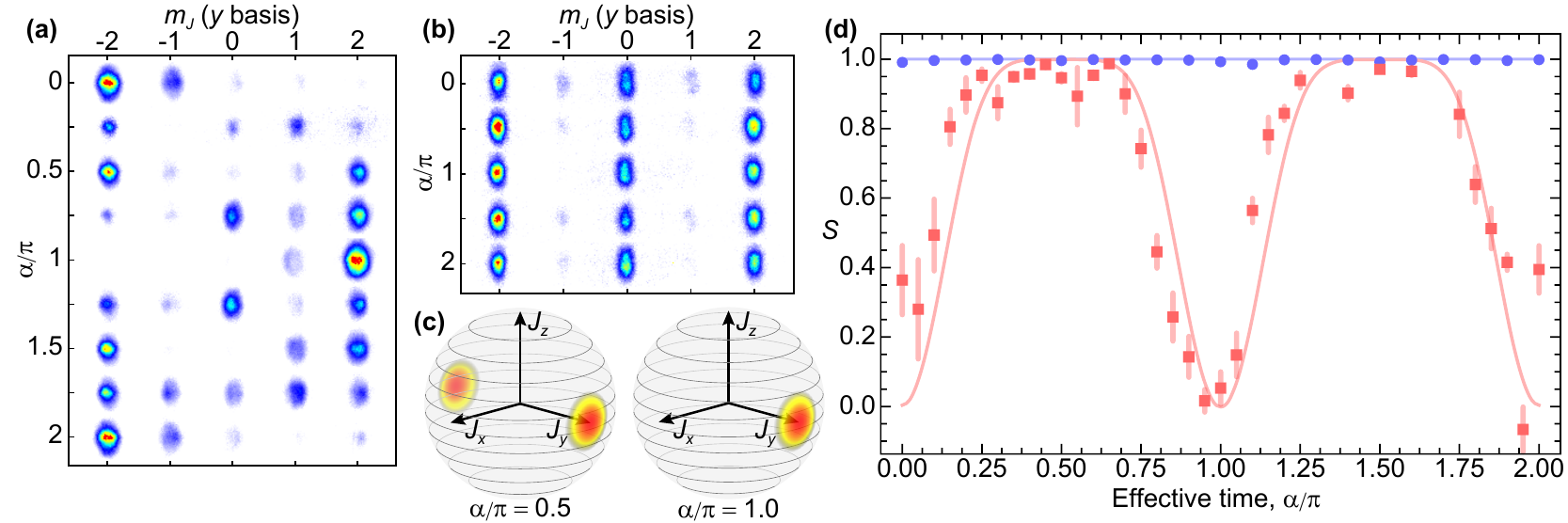}%
\caption{Squeezing of the artificial spin. Absorption images of the $y$-basis spin projection as a function of the effective squeezing time $\alpha$ when starting in \textbf{(a)} \css{\pi/2}{-\pi/2} and \textbf{(b)}  \jsn{2}{0}. \textbf{(c)} Density distributions for initial state \css{\pi/2}{-\pi/2} shown at effective time $\alpha/\pi = 0.5$ (left) and $\alpha/\pi = 1.0$ (right). \textbf{(d)} Linear entropy $S$ versus the effective squeezing time $\alpha$. The red squares and simulation line are for initial state \css{\pi/2}{-\pi/2} and the blue dots and simulation line are for \jsn{2}{0}. All error bars are one standard error of the mean.}%
\label{squeezingfig}%
\end{figure*}

\subsection{Nonlinear kick operation.} To realize the kicked top model, we additionally need to implement a nonlinear $\jzsq$ kick. Such a nonlinear spin operation gives rise to squeezing in the context of collective spin states~\cite{kitagawa-spinsqueezing}, such as in multi-mode condensates with mode-dependent interactions~\cite{Oberthaler-Chaos-2017} or through the collective, long-ranged interactions of many ions~\cite{Bollinger-literallyhundreds} or atoms in optical cavities~\cite{swingle-otocf}.

In experiments that are directly based on single spin-$J$ particles, the $\jzsq$ kick term relates instead to engineering a quadratic, $m_J$-dependent phase shift to the $z$-basis magnetic sublevels, creating nontrivial phase differences between neighboring $m_J$ states that impact their further evolution under subsequent linear rotations. For the case of emulating an artificial spin within a synthetic lattice of states, such a $\jzsq$ kick can be created through application of a quadratic potential of the site-energies in the absence of tunneling.

Alternatively, we directly engineer effective instantaneous relative phases at the different $m_J$ sites. This is accomplished by suddenly shifting the tunneling phase between two neighboring $m_J$ states to reflect the phase difference acquired during the instantaneous $\jzsq$ kick. As a concrete example for $J=2$, a $\jzsq$ kick with $\kappa = \pi/8$ leads to a relative phase accrual of $3\pi/8$ between the states $m_J=1$ and $m_J=2$. In our system, this phase difference is implemented by instantaneously shifting the phase of the $m_J=1\rightarrow m_J=2$ tunneling link as $t_1(\phi_1) \rightarrow t_1(\phi_1 +3\pi/8)$, or more generally $\phi_{n} \rightarrow \phi_{n} + (2n+1)\kappa$ for the $n \rightarrow n+1$ tunneling phase.\newpage

\section{Results}

\subsection{Squeezing the artificial spin.} We first examine the dynamics of our artificial spin under evolution governed by an effective spin-squeezing Hamiltonian $H_{\textrm{sq}}=\alpha_0\jzsq$. For any initial state, the $m_J$ population distribution will be unaffected in the $z$ basis. Therefore to explore the influence of the effective squeezing operation, we measure the $x$ and $y$ spin distributions by rotating into these measurement bases. The phase accrual of the $z$-basis $m_J$ states is accounted for by an appropriate modification of the phase terms of the various tunneling elements used to rotate the spins for measurement of $J_x$ and $J_y$.

For certain initial CSSs, evolution under $H_{\textrm{sq}}$  leads to the generation of correlations in the uncertainty of the spin value along the $x$, $y$, and $z$ directions. With increasing evolution time, the collective spin distribution undergoes periodic cycles of becoming squeezed and then returning to a simple, separable CSS. To characterize this behavior, we directly measure the spin distributions along the different spin directions $J_x$, $J_y$, and $J_z$. We combine these measurements to determine the linear entropy
\begin{equation}
S = 1 - \frac{\langle J_x \rangle^2+\langle J_y \rangle^2+\langle J_z \rangle^2}{J^2}
\end{equation}
of our artificial spin. In a CSS composed of many spin-$1/2$ particles, all the spins are aligned such that the length of the collective spin vector is $J$, yielding zero linear entropy. However, when the spin becomes maximally squeezed the net length of the spin vector becomes zero and $S$ takes on a value of one. In Fig.~\pref{squeezingfig}{d} we show the dependence of the linear entropy $S$ with increasing effective evolution time $\tau$, i.e. as the parameter $\alpha\equiv \alpha_0 \tau/\hbar$ increases. The measurements were carried out for two different initial states: the CSS \css{\pi/2}{-\pi/2} and the pure state \jsn{2}{0}.

The $y$-basis spin dynamics of the initial CSS are shown in Fig.~\pref{squeezingfig}{a}. Initially ($\alpha = 0$) aligned along the $-y$ axis, the CSS (red squares in Fig.~\pref{squeezingfig}{d}) should have a vanishing linear entropy. In experiment, imperfections in the state preparation and measurement rotations cause non-zero measurements at $\alpha=0$. At a larger effective evolution time ($\alpha = \pi/2$), the spin has rearranged itself such that half of the probability density is concentrated on each of the $-y$ and $+y$ axes (Fig.~\pref{squeezingfig}{c}, left) resulting in maximum linear entropy. Later, at $\alpha = \pi$, the spin realigns along the $+y$ axis and forms the zero entropy state \css{\pi/2}{+\pi/2}, as depicted in Fig.~\pref{squeezingfig}{c}. This process is also demonstrated in the $y$-basis absorption images shown in Fig.~\pref{squeezingfig}{a}.

In contrast to these entangling-disentangling dynamics, the pure angular momentum state \jsn{2}{0} is entirely unaffected by the squeezing operation, as by definition this state can support no important relative phase structure. This independence is illustrated by the data shown in Fig.~\pref{squeezingfig}{b} where the $y$-basis absorption images reflect no change across the entire spectrum of $\alpha$. Likewise, as seen in Fig.~\pref{squeezingfig}{d} (blue dots), the linear entropy of this non-CSS remains fixed at $S=1$ for all values of the effective evolution time $\alpha$.

While the initial CSS and non-CSS show wildly disparate dynamical behavior in their linear entropy under the squeezing Hamiltonian, they surprisingly behave similarly when considering instead the evolution of their out-of-time-ordered correlation functions (OTOCFs)~\cite{OTOCF}. These functions have been proposed as a suitable measure of dynamical entanglement and the scrambling of information in complex, many-body systems~\cite{scrambling2,scrambling1,swingle-otocf}, possibly even serving as a probe of many-body localization in disordered systems with interactions~\cite{OTOC-Zhai-MBL,Eduardo-MBL-OTOC}. Recently OTOCFs have been measured in a quantum spin system of interacting ions under a squeezing Hamiltonian~\cite{Bollinger-IonOTOCF} and in a chaotic spin chain based on nuclear magnetic resonance quantum simulation~\cite{Du-NMR-OTOCF}. Here, we use the wide tunability of our synthetic lattice parameters to measure OTOCFs for the first time with an atomic quantum gas. In particular, we demonstrate the suitability of this measure for tracking complex evolution of non-CSSs.

\begin{figure}%
\includegraphics[width=243pt]{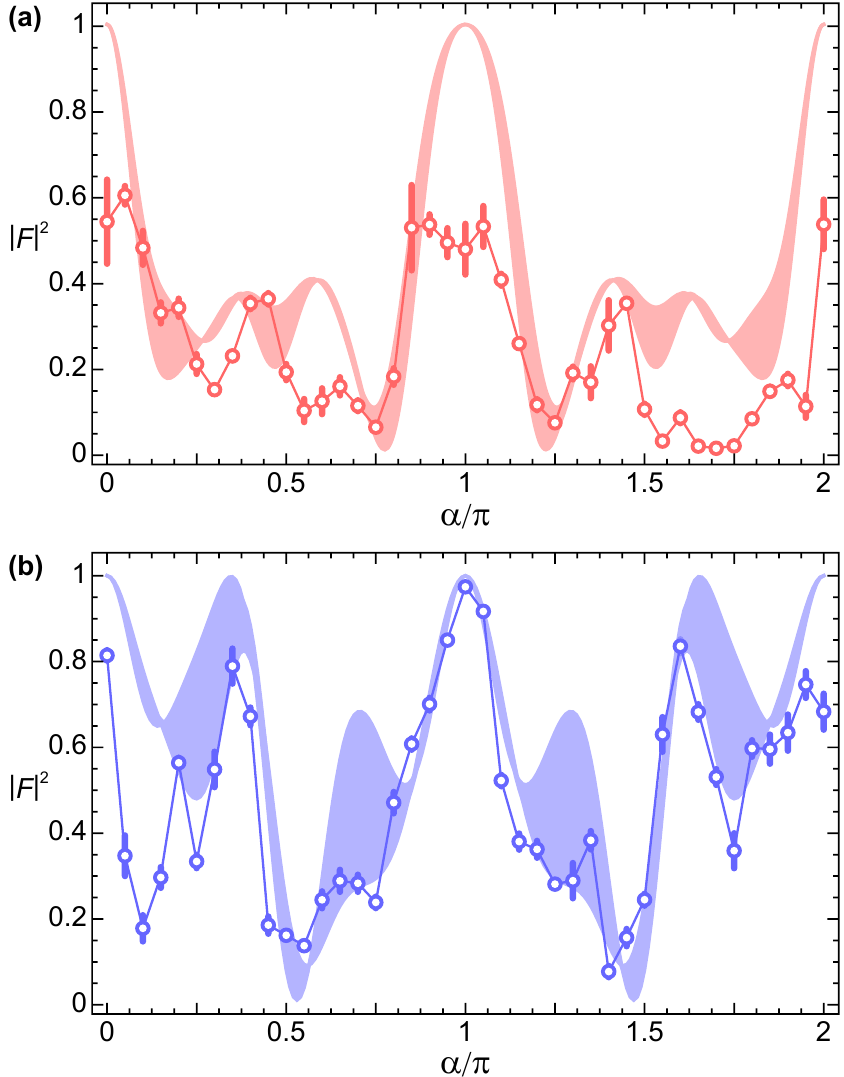}%
\caption{Out-of-time-ordered correlation function. $|F|^2$ for initial states \textbf{(a)} \css{\pi/2}{-\pi/2} and \textbf{(b)} \jsn{2}{0} as a function of the effective squeezing time $\alpha$. Shaded regions indicate the results of numerical simulations incorporating the uncertainty in the calibrated tunneling rate. All error bars are one standard error of the mean.}%
\label{OTOCFfig}%
\end{figure}

Essentially, OTOCFs probe the overlap between an initial state and that same state after some complex evolution characterized by a series of forward- and reverse-time operations. Following the terminology of Ref.~\cite{swingle-otocf}, we define the OTOCF as
\begin{align}
F(\alpha) &= \langle W_\alpha^\dagger V^\dagger W_\alpha V \rangle \ , \\
\intertext{where}
W_\alpha &= U(-\alpha)W U(\alpha) \\
\intertext{and}
U(\alpha) &= e^{-i \alpha \jzsq} \ ,
\end{align}
for commuting observables $W$ and $V$, which we set to be $W = V = e^{-i \frac{\pi}{4 \hbar} J_x}$. We perform the $\jzsq$ operations with an effective evolution parameter $\alpha$ as described above. Each of the $V$ and $W_\alpha$ operations involves tunneling for a time equivalent to a $\pi/4$ rotation, such that the full experimental duration (ignoring state preparation and readout) is equivalent to that of a global $\pi$ pulse. For a given initial state $|\Psi\rangle$, we measure $|F(\alpha)|^2$ by first applying the operator $F(\alpha)$ (by stepwise Hamiltonian evolution realizing the operators $V$, $W_\alpha$, $V^\dagger$, and $W_\alpha^\dagger$), then rotating to a measurement basis in which $|\Psi\rangle$ is an eigenstate, and finally determining the fraction of atoms that overlap with the initial state $|\Psi\rangle$. The numerical value of the OTOCF will generally be near one if simple, regular dynamics occur (perfect overlap $|F (\alpha = 0)|^2 = 1$ if there is no dynamical evolution) and zero if complex dynamics take place (somewhat opposite to the behavior of the linear entropy $S$).

In Fig.~\ref{OTOCFfig} we measure the OTOCF under evolution of our squeezing Hamiltonian for the same two initial states discussed previously: \css{\pi/2}{-\pi/2} and \jsn{2}{0}. In the case of an initial CSS (Fig.~\pref{OTOCFfig}{a}), the effective squeezing dynamics reflect those seen in the linear entropy, with $|F(\alpha)|^2$ taking a maximum value at $\alpha/\pi = \{0,1,2\}$. For an initial non-CSS, however, while the linear entropy was completely invariant as a function of $\alpha$, the OTOCF measurement in Fig.~\pref{OTOCFfig}{b} shows complex nontrivial dynamics. Thus, the OTOCF serves as a suitable probe for complex dynamics of the underlying Hamiltonian for more general initial states. We note that we find much better agreement between measurement and theory in the case of the initial state \jsn{2}{0}. This is a consequence of the fact that no additional experimental time is required for either state preparation or measurement rotation, reducing the susceptibility to experimental sources of error.

\begin{figure*}%
\includegraphics[width=400pt]{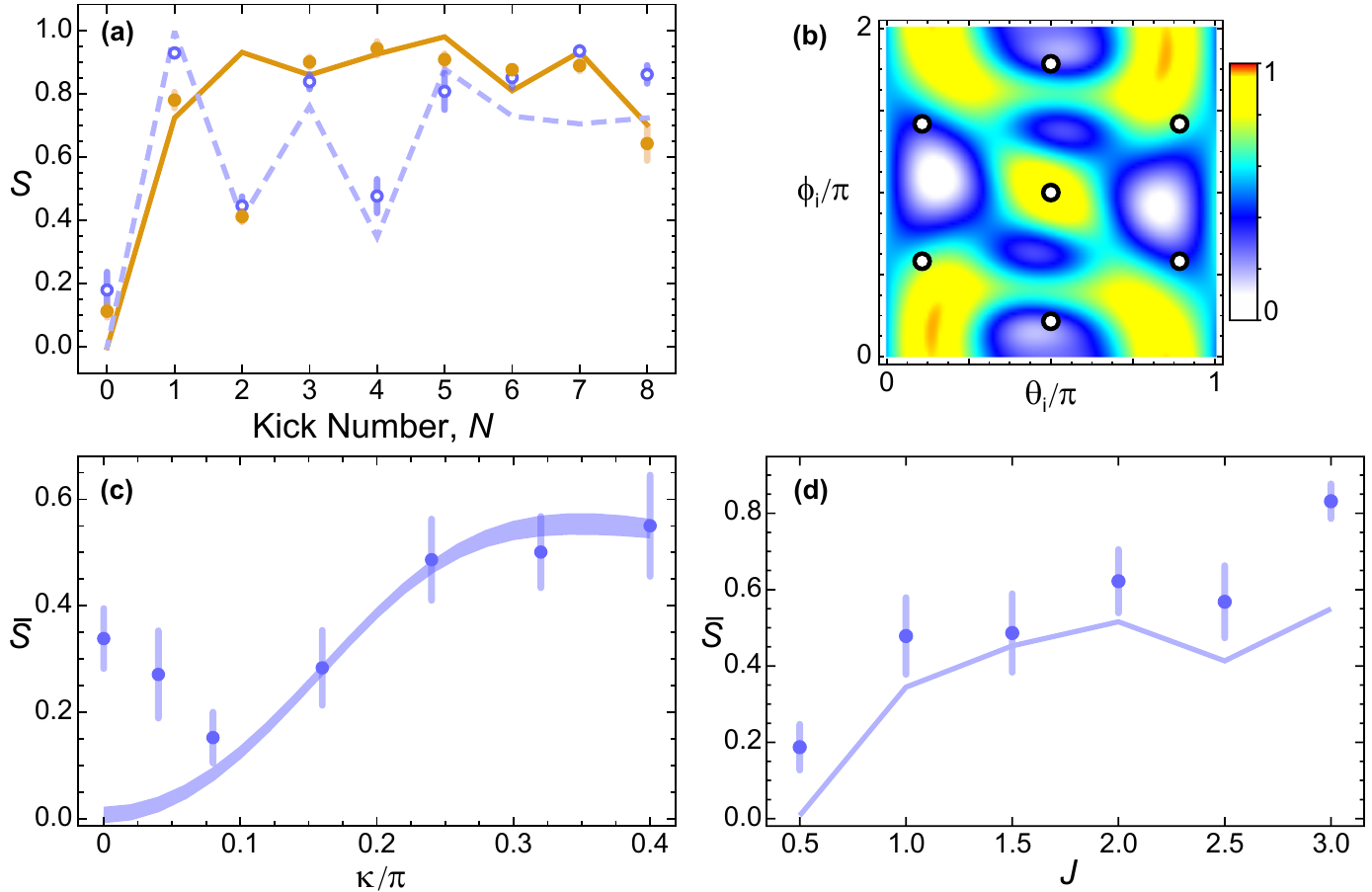}%
\caption{Chaotic behavior in the kicked-top model. \textbf{(a)} Linear entropy $S$ for initial state \css{\pi/2}{-\pi/2} measured after each kick in a set of eight kicks. The open blue dots and dashed blue simulation line are $(\rho, \kappa/2J) = (\pi/8, \pi/5)$ and the closed orange dots and solid orange simulation line are $(\rho, \kappa/2J) = (\pi/8, \pi/2)$. \textbf{(b)} Simulated linear entropy of a collective spin for different initial states. The color represents linear entropy averaged over $N\in\{5,6\}$ kicks, with respect to the color bar at right. The seven open black dots represent the measurements taken to calculate the averaged linear entropy $\bar{S}$. \textbf{(c)} $\bar{S}$ as a function of the kick strength $\kappa$ for $(\rho, J) = (\pi/8, 2)$. Shaded regions indicate results from a numerical simulation incorporating the uncertainty in the calibrated tunneling rate. \textbf{(d)} $\bar{S}$ as a function of the size of the spin $J$ for $(\rho, \kappa/2J) = (\pi/8, \pi/2)$.  Solid line connects points obtained from a numerical simulation. All error bars are one standard error of the mean.}%
\label{AvgSvsKappafig}%
\end{figure*}

\subsection{Chaotic behavior in the kicked top model.} Having demonstrated all of the necessary ingredients to simulate kicked tops with our artificial spins, we now engineer the full kicked top model and use it to explore unique aspects of chaotic behavior in a well controlled quantum system. For different initial CSSs and spin sizes $J$, we study the spin's linear entropy following evolution under Eq.~\ref{KTHam}. In Fig.~\pref{AvgSvsKappafig}{a}, for a spin size $J = 2$ and the initial state \css{\pi/2}{-\pi/2}, we study the dynamics of the linear entropy as a function of the number of applied kicks. Evolution under two different sets of kicked top parameters are shown: the filled orange circles relate to $(\rho,\kappa/2J) = (\pi/8, \pi/5)$ and the open blue circles relate to $(\rho,\kappa/2J) = (\pi/8, \pi/2)$. In both cases, the linear entropy almost immediately increases to near maximum after a single kick, showing the chaotic nature of the system under these conditions.

Our realization of the quantum kicked top model allows us to access the complete range of nonlinear coupling strengths with no deleterious side effects. This is in contrast to studies with cesium atoms~\cite{Jessen-csspin} and with superconducting qubits~\cite{Neill-Martinis-Chaos}, where only limited ranges of kick strength were explored. Using this full control of $\kappa$, we explore the onset of chaotic behavior as the nonlinear coupling strength $\kappa$ is increased. Because the presence of chaotic behavior in the system is very sensitive to the initial state, and because the classical phase-space boundaries (in terms of $\phi$ and $\theta$) between stable islands and chaotic regions change with increasing $\kappa$, we seek to reconstruct a global picture of how a typical initial state would evolve under given kicked top parameters. As such, we sample seven representative initial CSSs \css{\theta_i}{\phi_i} spread throughout phase space (illustrated in Fig.~\pref{AvgSvsKappafig}{b}), and measure the linear entropy averaged over these different cases. Moreover, to account for the fact that the dynamics of $S$ for a given orbit do not necessarily reach some constant value independent of the kick number, but in general undergo a complex evolution, we additionally average over the measured entropy $S$ for five and six kicks. The averaged (over initial state and kick number) linear entropy $\bar{S}$ is plotted as a function of nonlinear coupling strength $\kappa$ in Fig.~\pref{AvgSvsKappafig}{c}. A general agreement with the theoretical prediction (solid line) is observed, with a steady rise towards a larger linear entropy for increasing $\kappa$, signaling the onset of chaotic behavior. For small values of $\kappa$ the discrepancy between the theory and the data may be due to the lack of tunneling stability in our system, leading to an accumulation of error following many kick periods, state preparation, and state readout.

Finally, we use our unique ability to tune the size of our artificial spin to explore the initial crossover from the fully quantum regime towards the onset of classically chaotic behavior. For increasing $J$ values, where the initial CSSs become more and more sharply defined in terms of their $J_x$, $J_y$, and $J_z$ expectation values (normalized to $J$), one expects to reach a point where classical-like sensitivity to initial conditions can manifest even in quantized systems. A general correspondence between the onset of classical chaos and the development of high entanglement entropy in a quantum system has been observed for systems as small as $J = 3/2$~\cite{Neill-Martinis-Chaos}. Likewise, in the related chaotic system of kicked rotors, classical diffusive behavior has been observed for quantum systems of just two interacting rotors~\cite{Gadway-QtoC}. In Fig.~\pref{AvgSvsKappafig}{d}, we look at the growth of the averaged linear entropy $\bar{S}$ for a wide range of $J$ values from $1/2$ to $3$, for the case of $(\rho,\kappa/2J) = (\pi/8, \pi/2)$. For the smallest case of $J = 1/2$, which relates to just a single spin-$1/2$ particle, the spin should remain in a pure state with zero linear entropy at all times and for all initial states. As the system size grows, however, theoretical calculations (solid line) predict a steady trend towards increasing averaged linear entropy, signaling a crossover to increasingly classical-like chaotic behavior. We indeed observe a similar trend in the dynamical evolution of our artificial spins, with mostly regular evolution for small $J$ giving way to significantly more entropy generation for larger $J$.

\section{Discussion}

Our study based on Hamiltonian engineering in a synthetic lattice offers a new approach to exploring the correspondence between quantum and classical dynamics, offering the possibility of tuning the size of a driven synthetic spin. Here, we have been limited to exploring only modest values of $J$, due to the increasing duration required for rotations of the effective spin for increasing $J$ values. However, straightforward improvements to our experiment should allow us to probe signatures of chaos in artificial spins of size $J \sim 10 - 20$. Currently, we are limited primarily by the spatial separation of the wavepackets relating to the many discrete momentum orders. This loss of near-field coherence may be mitigated in the future, however, by creating more spatially extended condensates, or through refocusing (echo) protocols.

Our demonstration of a synthetic lattice approach to kicked top studies also suggests that related platforms, having similar levels of local and dynamical parameter control, could also be used to explore quantum chaos. In particular, the high degree of control in discrete photonic systems~\cite{Szameit-GlauberFock} should enable similar explorations, perhaps with extensions to much larger effective spin sizes.

In addition to the tunable size of our synthetic spins, the wide control afforded by synthetic lattice techniques should also enable further studies on the dynamics of modified kicked tops belonging to distinct symmetry classes~\cite{Kus1987}. More generally, synthetic lattices should even enable the precise implementation of random unitary operations. This raises the interesting prospect of exploring boson sampling problems~\cite{BosonSamplingReview} with few-particle states in synthetic lattices.

Lastly, we remark on the influence of atomic interactions on the dynamics in our synthetic kicked top. Under present experimental conditions, the tunneling energy $t$ dominates heavily over the mean-field interaction strength of our condensate atoms $\mu$ (with $t/\mu \gtrsim 5$), such that we do not expect any substantial modification of the dynamics as compared to non-interacting particles. However, by working at smaller values of $t$, we can enter the regime where interactions lead to correlated dynamics. That is, cold collisions give rise to an effective nonlinear interaction in the collective spin of many spin-$1/2$ particles~\cite{Oberthaler-Chaos-2017} (i.e. nonlinear interactions in a momentum-space double well~\cite{MSL-interactions}). The use of a synthetic spin, as compared to a real spin, also opens up the intriguing possibility of exploring the driven dynamics of a system of many collectively interacting large-$J$ particles.

\end{document}